\documentclass[acp, manuscript]{copernicus}

\usepackage{tikz}
\usetikzlibrary{patterns}
\usetikzlibrary{shapes}
\usetikzlibrary{decorations.text}
\usetikzlibrary{shapes.multipart}
\usetikzlibrary{arrows,shapes.geometric,positioning}
\usetikzlibrary{shapes,positioning,decorations.pathmorphing}
\usetikzlibrary{calc}
\usepackage{booktabs}
\usetikzlibrary{backgrounds}
\usepackage{varwidth}
\usetikzlibrary{fit}

\usepackage{amsmath}
\usepackage{latexsym,euscript,textcomp}
\usepackage{color,graphics,epsf,dcolumn}
\usepackage{epstopdf,ulem}
\usepackage{blindtext}
\usepackage{threeparttable}
\newcommand{\beq}{\begin{equation}}
\newcommand{\eeq}{\end{equation}}
\newcommand{\pt}{\partial}

\begin{document}
\nolinenumbers



\title{Re-appraisal of the global climatic role of natural forests for improved climate projections and policies}

\Author[1,2]{Anastassia M.}{Makarieva}
\Author[1]{Andrei V.}{Nefiodov}
\Author[3]{Anja}{Rammig}
\Author[4]{Antonio Donato}{Nobre}

\affil[1]{Theoretical Physics Division, Petersburg Nuclear Physics Institute, 188300 Gatchina, St.~Petersburg, Russia}
\affil[2]{Institute for Advanced Study, Technical University of Munich, Lichtenbergstra{\ss}e 2~a, 85748 Garching, Germany}
\affil[3]{Technical University of Munich, School of Life Sciences, Hans-Carl-von-Carlowitz-Platz 2,  85354 Freising, Germany}
\affil[4]{Centro de Ci\^{e}ncia do Sistema Terrestre INPE, S\~{a}o Jos\'{e} dos Campos, 12227-010 S\~{a}o Paulo, Brazil}


\runningtitle{Re-appraisal of the global climatic role of natural forests}

\runningauthor{Makarieva and Nefiodov}

\correspondence{A. M. Makarieva (ammakarieva@gmail.com)}

\received{}
\pubdiscuss{} 
\revised{}
\accepted{}
\published{}


\firstpage{1}

\maketitle

\begin{abstract}
Along with the accumulation of atmospheric carbon dioxide, the loss of primary forests and other natural ecosystems is a major disruption of the Earth system causing global concern. Quantifying planetary warming from carbon emissions, global climate models highlight natural forests{\textquoteright} high carbon storage potential supporting conservation policies. However, some model outcomes effectively deprioritize conservation of boreal and temperate forests suggesting that increased albedo upon deforestation could cool the planet. 
Potential conflict of global cooling versus regional forest conservation could harm environmental policies. Here we present theoretical and observational evidence to demonstrate that, compared to the carbon-related warming, the model skills for assessing climatic impacts of deforestation is low. We argue that deforestation-induced global cooling results from the models{\textquoteright} limited capacity to account for the global effect of cooling from evapotranspiration of intact forests. Transpiration of trees can change the greenhouse effect via small modifications of the vertical temperature profile. Due to their convective parameterization (which postulates a certain critical temperature profile), global climate models do not properly capture this effect. This parameterization may lead to underestimation of warming from the loss of evapotranspiration in both high and low latitidues,  and therefore, conclusions about deforestation-induced global cooling are not robust. To avoid deepening the environmental crisis, these conclusions should not inform policies of vegetation cover management. Studies are mounting quantifying the stabilizing impact of natural ecosystems evolved to maintain environmental homeostasis. Given the critical state and our limited understanding of both climate and ecosystems, an optimal policy would be a global moratorium on the exploitation of all natural forests.
\end{abstract}

\introduction  

\label{intr}

\begin{figure*}[!t]
\begin{minipage}[p]{1\textwidth}
\centering\includegraphics[width=0.5\textwidth,angle=0,clip]{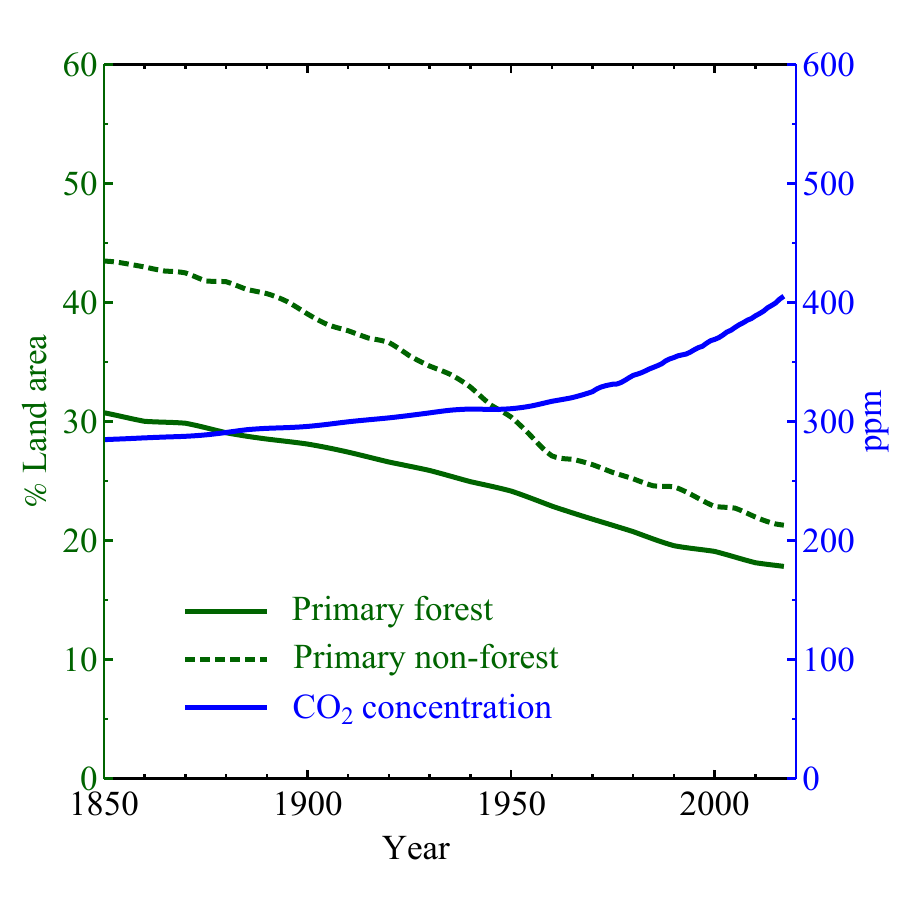}
\end{minipage}
\caption{
Decline of primary forest and non-forest ecosystems \citep[left axis, data from][their Fig.~7]{hurtt20} and increase of atmospheric CO$_2$ (right axis, data downloaded from \url{https://www.eea.europa.eu/data-and-maps/daviz/atmospheric-concentration-of-carbon-dioxide-5}). During the industrial era, primary ecosystems have declined, and CO$_2$ concentration has grown, by approximately one half.
}
\label{dist}
\end{figure*}

The Earth is suffering from climate destabilization and ecosystem degradation (Fig.~\ref{dist}), and the humanity seeks to stop both \citep{ipbes19}.
Policies for global climate stabilisation, with the focus on decarbonization, are informed by the outcomes of global climate models 
that formalize our evolving understanding of the Earth system -- currently, by the
model simulations from the 6th Coupled Model Intercomparison (CMIP6) for the 6th Assessment Report of the Intergovernmental Panel on Climate Change 
\citet[IPCC-AR6; IPCC][]{ipcc21}.
 With the Intergovernmental Science-Policy Platform on Biodiversity and Ecosystem Services (IPBES) formed twenty four years later than IPCC, the ecosystem preservation narrative is less formally developed \citep{wilhere21}. Proponents of ecosystem preservation often borrow from the decarbonization argumentation and invoke the carbon storage potential of natural forests as a major illustration of their climatic importance. For example, the ground-breaking proforestation initiative in the United States began with emphasizing how much carbon the unexploited forests can remove from the atmosphere if allowed to develop to their full ecological potential \citep{moomaw2019}.

However, the carbon-storage argument for temperate and boreal forests is undermined by the fact that global climate models
suggest that deforestation in these regions could cool the planet. Here increased albedo is estimated to overcome the
warming caused by deforestation-induced carbon emissions \citep[][Fig.~2.17]{ipcc19}, even if the latter
can be underestimated \citep{schepaschenko21}. These model outcomes have been known for quite a while \citep[e.g,][]{snyder04},
but recently these ideas gained prominence approaching implementation. A recent {\it Science} commentary warned that regrowing boreal forests  would not make the Earth cooler \citep{pearce22}, a conclusion that is purely derived from global climate model simulations \citep[e.g.,][]{hertog22}. The World Resources Institute{\textquoteright}s report {\textquotedblleft}Not just carbon{\textquotedblright} noted that the increased albedo from deforestation would cool the Earth  and emphasized that the positive climate role of boreal forests is only local \citep{seymour22a,seymour22b}. Accordingly, a recent study in {\it Nature Ecology and Evolution} did not include primary boreal forests  into Nature{\textquoteright}s critical assets \citep{chaplin-kramer22}. One of the criteria for an ecoregion to be classified as a critical asset, was its proximity to people -- and primary boreal forests are often distant from any human settlements (which is a major reason for why they are still primary). Together, these mainstream messages not only de-emphasize the preservation of natural boreal and, to a lesser degree, temperate forests, but implicitly incentivize their destruction.

In this Perspective, we would like to ring an alarm bell by showing 
that this potentially biased picture of the role of natural forests, in particular boreal forests, for stabilizing Earth's climate is based on a few model assumptions ruling out
important evapotranspiration feedbacks and can result in policies deepening rather than mitigating the climate crisis.  We also outline a possible path forward.

\section{Global cooling from plant transpiration}

\subsection{Local versus global cooling}

\begin{figure*}[!t]
\begin{minipage}[p]{1\textwidth}
\centering\includegraphics[width=0.7\textwidth,angle=0,clip]{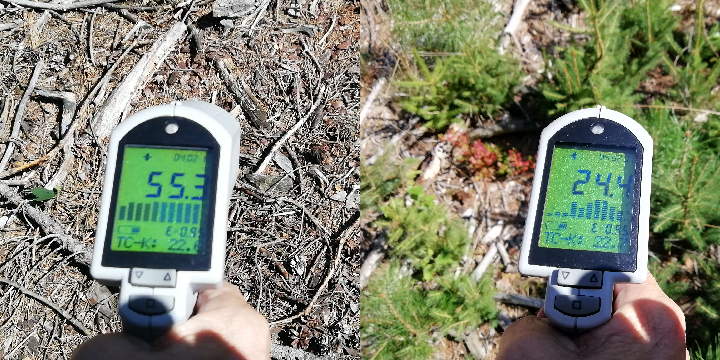}
\end{minipage}
\caption{
Local cooling from plant transpiration. With incoming solar radiation of about $1$~kW~m$^{-2}$, dry
area on the deforested plot (left panel) has temperature of $55.3$~\degree{C}. Young transpiring trees (right panel)
lower the surface temperature by almost 30~\degree{C}. Distance between the two spots is 1 meter.
Measurements and photo credit Jan Pokorn{\'y}.
}
\label{loc}
\end{figure*}

We argue that the conclusion of a cooler Earth upon the loss of boreal forests stems from the limited
capacity of global climate models to quantify another effect of the opposite sign: global cooling from forest transpiration.
That transpiring plants provide local cooling is well-known \citep[e.g.,][and see Fig.~\ref{loc}]{hurina16,alkama16,ellison2017,hesslerova18}. Instead of converting to heat, a certain part of absorbed solar energy is spent to break the intermolecular (hydrogen) bonds between the water molecules during evapotranspiration. As a result, the evaporating surface cools.

When more sunlight is reflected back to space, the planet receives less energy and it is intuitively clear that it must cool.
In comparison, although evaporation does cool locally, the captured energy does not disappear but is released upon condensation elsewhere in the Earth system. In contrast with the well developed methodology of explaining the rising planetary temperature with increasing CO$_2$ \citep[e.g.,][and references therein]{benestad17}, how  and whether loss of plant transpiration could warm the planet remains unclear. While the IPCC science does recognize that global cooling from plant transpiration exists \citep[][Fig.~2.17]{ipcc19}, its description is not to be found in textbooks. However, with the environmental science being inherently transdisciplinary, understanding this effect is important for the broader community of ecosystem researchers and conservationists,  as it will enable a critical assessment of model outputs offered to guide large-scale vegetation management.

\subsection{Conceptual picture}

\begin{figure*}[!t]
\begin{minipage}[p]{1\textwidth}
\centering\includegraphics[width=0.95\textwidth,angle=0,clip]{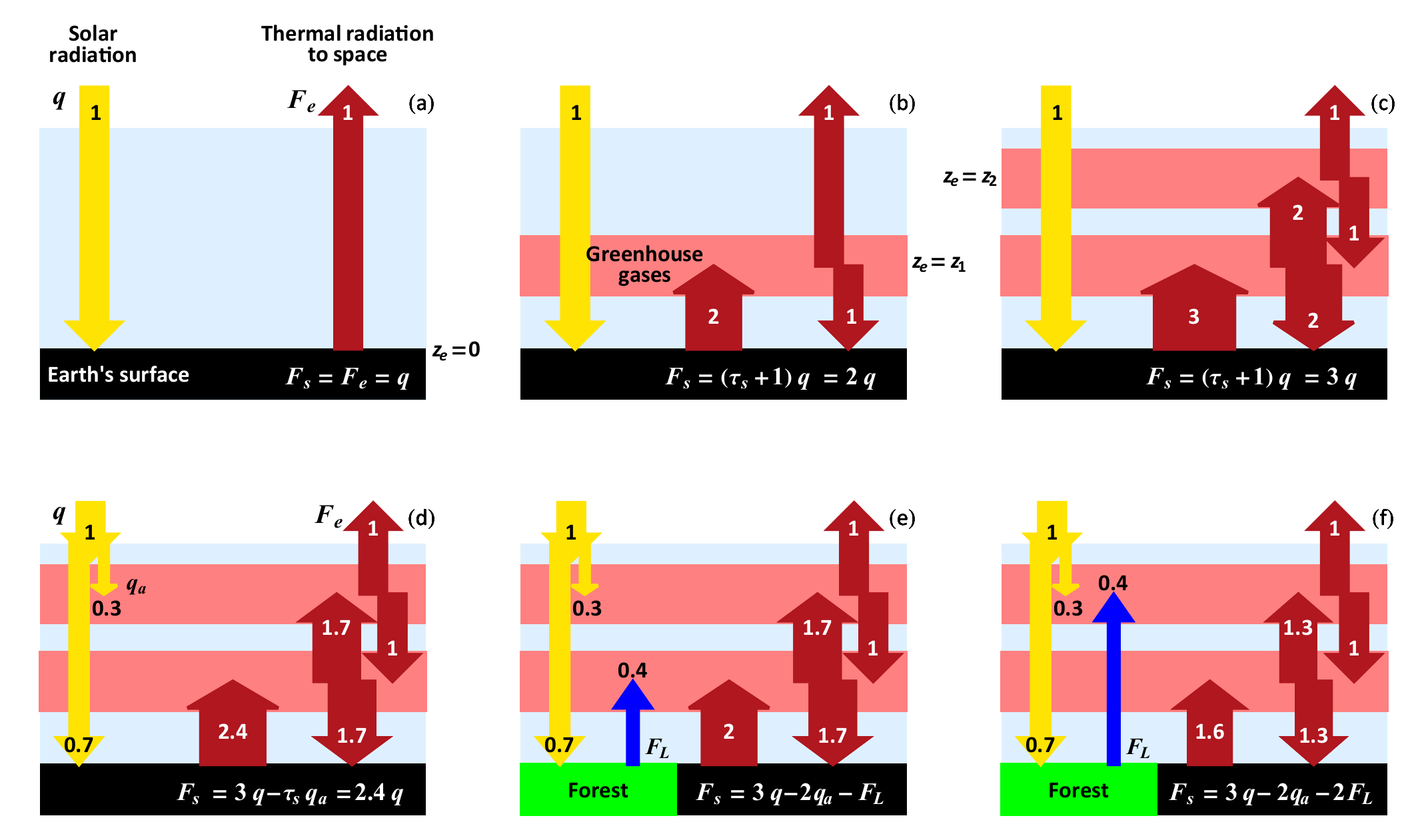}
\end{minipage}
\caption{
Scheme to illustrate the dependence of the planetary surface temperature on the amount of greenhouse substances (a-c)
and on the magnitude and spatial distribution of the non-radiative energy fluxes (d-f).  Thickness of each layer of the greenhouse substances corresponds to unit optical depth $\tau=1$ (one free path of thermal photons -- the mean distance between two consecutive acts of absorption and re-emission by the absorber molecules); $\tau_s$ is the total number of layers: $\tau_s =0$ in (a), $1$ in (b) and $2$ in (c-f). A {\textquotedblleft}gray{\textquotedblright} atmosphere is assumed, where absorption of thermal radiation is the same for all wavelengths \citep{ramanathan78,makarieva01,gorshkov02}. Thermal radiation of the planetary surface $F_s = \sigma T_s^4$ (W~m$^{-2}$) and of the upper radiative layer to space $F_e = \sigma T_e^4$  are related to surface temperature $T_s$ and temperature of the upper radiative layer $T_e$ by Stefan-Boltzmann law, where $\sigma =5.7\times 10^{-8}$~W~m$^{-2}$~K$^{-4}$  is the Stefan-Boltzmann constant. All energy fluxes are shown in the units of absorbed solar radiation $q$, which is in the steady-state equal to thermal radiation emitted by the planet ($F_e = q$);  in (d-f), $q_a=0.3q$ is solar energy absorbed by the atmosphere;  in (e,f), $F_L$ is the non-radiative heat flux accounting for both sensible and latent heat. With $q = 239$~W~m$^{-2}$, configuration (f) approximately corresponds to the modern Earth with $q_a = 78~{\rm W~m}^{-2} \simeq 0.3 q$, $F_s = 390 ~{\rm W~m}^{-2} \simeq 1.6 q$, $F_L = 97 ~{\rm W~m}^{-2} \simeq 0.4q$ \citep{trenberth09}. Thermal radiation is emitted to space from mean height $z_e$: $z_e = 0$ in (a), $z_e = z_1 > 0$ in (b) and $z_e = z_2>z_1$ in (c-f).
}
\label{ghe}
\end{figure*}

To illustrate the effect, we will use a simple model of energy transfer (Fig.~\ref{ghe}).
The greenhouse substances are represented by discrete layers that absorb all incoming thermal radiation and radiate
all absorbed energy equally up and down. In the absence of absorbers, the Earth{\textquoteright}s surface emits as much thermal radiation as it receives from the Sun (Fig.~\ref{ghe}a).  Each layer of the greenhouse substances redirects part of the thermal radiation back to the Earth{\textquoteright}s surface. As a result, the planetary surface warms, and the more so, the greater the amount of absorbers (cf. Fig.~\ref{ghe}b and c).

When a certain part of the incoming solar radiation is absorbed in the upper atmosphere (for example, by aerosols or clouds),
it escapes interaction with the absorbers beneath. Accordingly, the planetary surface cools by an amount by which the absorbers would
multiply this escaping part if it dissipated to thermal radiation at the surface (cf. Fig.~\ref{ghe}c and d). This illustrates how 
{\it where the solar energy dissipates to thermal radiation}, with unchanged amount of greenhouse gases and total absorbed solar energy, impacts the planetary surface temperature.

Similarly, in the presence of the non-radiative heat fluxes of sensible and latent heat, the amount of solar energy converted to thermal radiation  at the surface diminishes -- and so does the amount of thermal radiation redirected by the absorbers back to the surface. 
Surface thermal radiation and temperature decline (cf. Fig.~\ref{ghe}d and e,f). The non-radiative fluxes {\textquotedblleft}hide{\textquotedblright} a certain part of absorbed solar energy from the greenhouse substances easing its ultimate release to space. 
Convection, condensation and precipitation {\textquotedblleft}deposit the latent heat removed from the surface above the level of the main water vapor absorbers, whence it is radiated to space{\textquotedblright} \citep{bates03}. This energy escaping partially from interaction with the absorbers is the cause of global cooling from plant transpiration.

A related process is the atmospheric transport of heat from the equator to higher latitudes, where the water vapor concentration
in the colder atmosphere is smaller. This transport likewise {\textquotedblleft}hides{\textquotedblright} a certain part
of solar energy absorbed at the equator from the abundant greenhouse substances (water vapor) in the warm tropical atmosphere. In the result, despite the amount of absorbers does not change, the globally averaged greenhouse effect diminishes and the planetary surface cools \citep{bates99,caballero01}. \citet[][]{marvel13} modeled an idealized atmosphere with two strong circulation cells connecting the equator and the poles. With such a circulation, the Earth{\textquoteright}s surface became eleven degrees Kelvin cooler than the modern Earth \citep[][their Fig.~1e and Fig.~3 bottom]{marvel13}.

Increasing the non-radiative flux (from zero in (d) to $F_L>0$ in (e) and (f)) decreases surface thermal radiation $F_s$ by a magnitude proportional to $F_L$ itself and to the number of absorbing layers $\Delta \tau$ beneath the height where this flux dissipates to thermal radiation ($1$ in (e) and $2$ in (f)). Historical deforestation affected about $13\%$ of land area (or $3.8\%$ of planetary surface) (Fig.~\ref{dist}). With the global mean latent flux of $F_L = 80$~W~m$^{-2}$, if deforestation has reduced this flux by thirty per cent ($\Delta F_L \sim -0.3 F_L$), this could increase the surface radiation by $-0.038 \Delta F_L \sim 0.9$~W~m$^{-2}$ (cf. Fig.~\ref{ghe}d and e) or twice that value (cf. Fig.~\ref{ghe}d and f), Table~\ref{tab1}. Given an equilibrium climate sensitivity $\varepsilon \sim 1 $~K/(W~m$^{-2}$) \citep{zelinka20}, the latter case corresponds to a warming of about two degrees Kelvin (Table~\ref{tab1}). This should be manifested as an increase in the temperature difference between the surface and the upper radiative layer (the mean temperature lapse rate $\overline{\Gamma} = (T_s - T_e)/z_e$, Fig.~\ref{ghe}). If the optical thickness of the atmosphere is greater, the cooling will be proportionally larger.

\begin{table}[!ht]
     \caption{Estimates of global warming from the loss of tree transpiration associated with deforestation; $\Delta F_L$~($\%$ of the global mean value $F_L$) is the local reduction of latent heat on the deforested area $\Delta S$~($\%$ of land area $S_{\textrm{l}}$); $\Delta T$ is the change of {\it global} surface temperature upon deforestation.}\label{tab1}
    \begin{threeparttable}
    \footnotesize
    \centering    
    \begin{tabular}{llllll}
    \hline \hline 
    Study & Area affected&  $\Delta S$ ($\%$) & $\Delta F_L$ ($\%$) & $\Delta T$ (K) & \\
            
    \hline
\citet{snyder04}    &	Tropical$^{\rm *}$ 	&	$16$	&	$-30$	&	$0.24$	&	\\ [2ex]
\citet{davin10}        &	Global$^{\rm **}$ 	&	$90$	&		&	$0.55$	&	\\ [2ex]
This work  (Fig.~\ref{ghe}d,f)      &	Historical$^{\rm ***}$	&	$13$	&	$-30$	&	$2$	&	\\ [2ex]
\hline   \hline
    \end{tabular}
\begin{tablenotes}[para,flushleft]
$^{\rm *}$ Tropical forests replaced by deserts in a coupled atmosphere-biosphere model.\newline
$^{\rm **}$ Deforestation of a fully forested planet without changing the albedo; $\Delta T$ 
is the sum of two effects, change in roughness and change in evapotranspiration efficiency 
as shown in Table~1 of \citet{davin10}.\newline
$^{\rm ***}$ Estimated as $\Delta T \sim -0.29 (\Delta S/S_{\textrm{l}}) \Delta F_L \Delta \tau \varepsilon$, 
assuming that deforestation reduces latent heat flux by $30\%$ of $F_L=80$~W~m$^{-2}$ \citep{trenberth09}  on $13\%$ of land (the area affected by historical deforestation (Fig.~\ref{dist}), $0.29$ is the relative global land area)  with $\Delta \tau = \tau_s = 2$ as optical depth of the atmosphere (Fig.~\ref{ghe}d,f);
$\varepsilon \sim 1 $~K/(W~m$^{-2}$) is the assumed equilibrium climate sensitivity to radiative forcing.
\end{tablenotes}
\end{threeparttable}
\end{table}

\subsection{Dependence of global transpirational cooling on atmospheric circulation}

The higher up convection transports heat, the more pronounced global cooling it exerts  as the energy is radiated more directly to space from the upper atmospheric layer (cf. Fig.~\ref{ghe}e and f).

Besides the altitude, it matters how rapidly the air ultimately descends. 
When the air rises and increases its potential energy in the gravitational field, its internal energy accordingly declines, and it cools.
As originally evaporation cooled the evaporating surface, the release of latent heat during condensation in the rising air partially offsets this decline of the internal energy of air molecules  making the air warmer than it would be without condensation. Radiating this extra thermal energy to space takes time. The more time the air warmed by latent heat release spends in the upper atmosphere (above the main absorbers), the more energy is radiated unimpeded to space and the stronger the global transpirational cooling.  With the characteristic radiative cooling rate of the order of $2$~K~day$^{-1}$, it takes about fifteen-thirty days to radiate the latent heat released by tropical moist convection \citep{goody03}.

\begin{figure*}[!t]
\begin{minipage}[p]{1\textwidth}
\centering\includegraphics[width=0.7\textwidth,angle=0,clip]{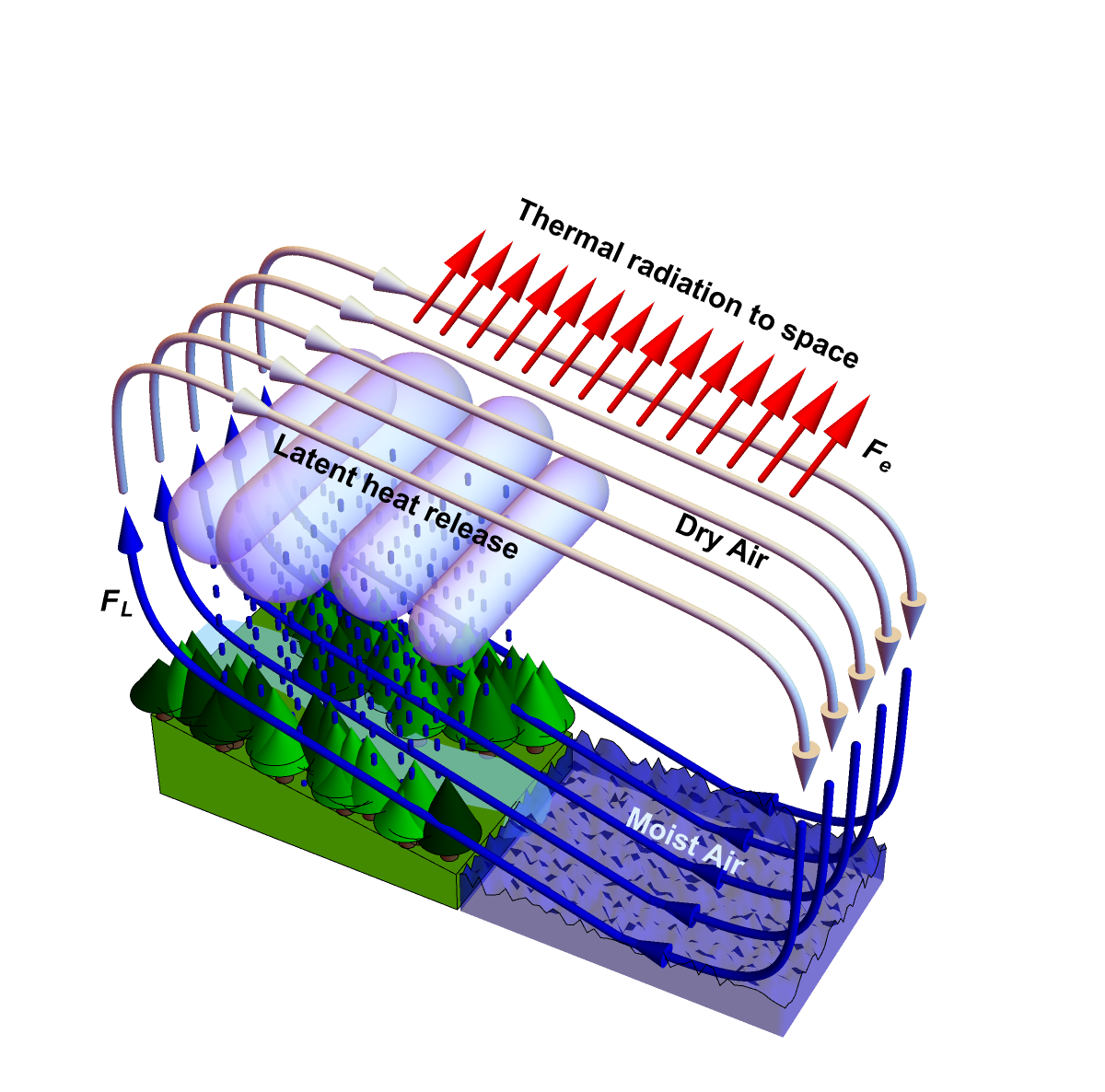}
\end{minipage}
\caption{
Visualization of the link between transpirational cooling and air circulation. After latent heat is released upon condensation, 
this energy can be radiated to space while the air travels back to the ocean in the upper troposhere.
}
\label{bp}
\end{figure*}

Therefore, the long-distance moisture transport \citep[including the biotic pump run by forests,][]{mg07} enhances global transpirational cooling: moist air travels for many days thousands of kilometers from the ocean to land where it ascends and latent heat is released.
Then the dry air warmed by latent heat makes the same long way back in the upper atmosphere radiating energy
to space (Fig.~\ref{bp}). If, on the contrary, the warmed air descends rapidly and locally, then most heat is brought back to the surface before it is radiated, and the net cooling effect can be nullified. Disruptions in the long-distance moisture transport (e.g., by deforestation) and violent local rains should warm the Earth. In smaller convective clouds up to a quarter of ascending air descends locally at a relatively high vertical velocity \citep{heus08,katzwinkel14}.

Global climate models do not correctly reproduce either the long-distance ocean-to-land moisture transport or the moisture transport over the ocean \citep{sohail22}. For example, the Amazon streamflow is underestimated by up to 50\% \citep[][their Fig.~5]{ma06,hagemann11}. This corresponds to a 10\% error in the global 
continental streamflow, the latter being of the same order as global continental evaporation.  Nor do global climate models correctly reproduce
how the local diurnal cycle of convection changes upon deforestation producing extreme low and high temperatures \citep[their Fig.~7]{lejeune17}. These are indirect indications of the models{\textquoteright} limited capacity to reproduce global transpirational cooling.

\subsection{Global transpirational cooling in global climate models}

We have seen that, for a given amount of absorbers, surface temperature is determined by the vertical distribution of the non-radiative heat fluxes (Fig.~\ref{ghe}d-f). But these fluxes themselves depend on the vertical temperature gradient: if the air temperature declines with height faster than a certain critical lapse rate\footnote{Lapse rate $\Gamma$ is the absolute magnitude of the vertical temperature gradient, $\Gamma \equiv -\pt T/\pt z$.}, the atmosphere is unstable to convection. The non-radiative heat fluxes originate proportional to the difference between the actual and the critical temperature lapse rates \citep{ramanathan78}.

Therefore, strictly speaking, it is not justified to freely vary where and how the non-radiative heat fluxes dissipate to thermal radiation, not paying attention to whether the resulting vertical temperature profile is consistent with their specified values. However, since the non-radiative (convective) and net radiative energy fluxes in the Earth{\textquoteright}s atmosphere are of the same order of magnitude \citep[$100$~and~$60$~W~m$^{-2}$, respectively][]{trenberth09}, a rough estimate of global transpirational cooling can be obtained from  considering the radiative transfer alone as done in Fig.~\ref{ghe}d-f. (This would not be possible if the convective fluxes were an order of magnitude higher than the radiative flux). We emphasize that our goal here is not  to obtain an accurate estimate of global transpirational cooling, but to present plausible arguments showing that it can be large.

An exact estimate of what happens when the evapotranspiration and the latent heat flux are suppressed on a certain part of land area requires solving the problem simultatenously for the radiative-convective transfer and the temperature profile. This problem is too complicated for modern global climate models, which therefore apply the so-called convective parameterization. The idea is to postulate the (generally unknown) value of a critical temperature lapse rate 
instead of solving for it. While the numerical simulation is run,  {\it {\textquotedblleft}whenever the radiative equilibrium lapse rate is greater than the critical
lapse rate, the lapse rate is set equal to the critical lapse rate{\textquotedblright}} \citep{ramanathan78}. Therefore, by construction, global climate models cannot 
provide any independent information about the climatic effect of evapotranspirational cooling -- that should be manifested as the change in the global mean lapse rate --
besides what was fed into them {\it a priori} via convective parameterization. 

Global climate models have been built with a major goal to assess radiative forcing from changing carbon dioxide concentrations.
They do have this capacity: this forcing can be approximately estimated assuming an unchanged atmospheric temperature profile. It is under this assumption that  \citet{arrhenius1896} obtained the first ever estimate of global warming from CO$_2$ doubling\footnote{If the lapse rate $\Gamma$ is known, an alternative way to calculate how surface temperature rises with increasing concentration of greenhouse substances is to calculate the change of radiative height $\Delta z_e = z_2 - z_1$ (cf. Fig.~\ref{ghe}b and c) using the hydrostatic equilibrium; then use 
$\Delta T_s = \Gamma \Delta z_e$.}. But radiative forcing caused by the suppression of evapotranspiration is a conceptually different problem for which convective parameterization precludes a solution that would be non-zero in the first order. Therefore, in the models, global warming resulting from the loss of transpirational cooling is, for the same deforested area, at least one order of magnitude smaller than our estimate (Table~\ref{tab1}).  For example, 
according to global climate models, tropical deforestation on $16\%$ of land area would produce a global warming of $0.2$~K \citep{snyder04}, while converting most land from forest to grassland (with unchanged albedo) would warm the Earth by about half a degree Kelvin \citep{davin10}, see Table~\ref{tab1}.

As an illustration of the lack of conceptual clarity with regard to global transpirational cooling, one can refer to the conclusion of \citet[][their Table~1]{davin10} that modeled global warming due to the loss of evapotranspiration is a {\textquotedblleft}non-radiative{\textquotedblright} forcing as compared to the change of albedo. This conclusion is reached by noting that loss of evapotranspiration practically does not change the radiation balance at the top of the atmosphere: $\Delta F_e \to 0$ such that $\Delta T/\Delta F_e \gg \varepsilon$. However, using this logic, CO$_2$ increase would not be a radiative forcing either, because, once the planetary temperature equilibrates, CO$_2$ increase {\it per se} (feedbacks absent) does not change the outgoing radiation at the top of the atmosphere. Indeed, Fig.~\ref{ghe} illustrates how the planetary temperature changes due to the radiative forcing from an increased amount
of greenhouse substances (a-c) and due to the radiative forcing from changing non-radiative fluxes (d-f). The incoming solar and outgoing longwave radiation remain the same in all cases ($F_e = q$ and $\Delta F_e = 0$).

\begin{figure*}[!t]
\begin{minipage}[p]{1\textwidth}
\centering\includegraphics[width=0.95\textwidth,angle=0,clip]{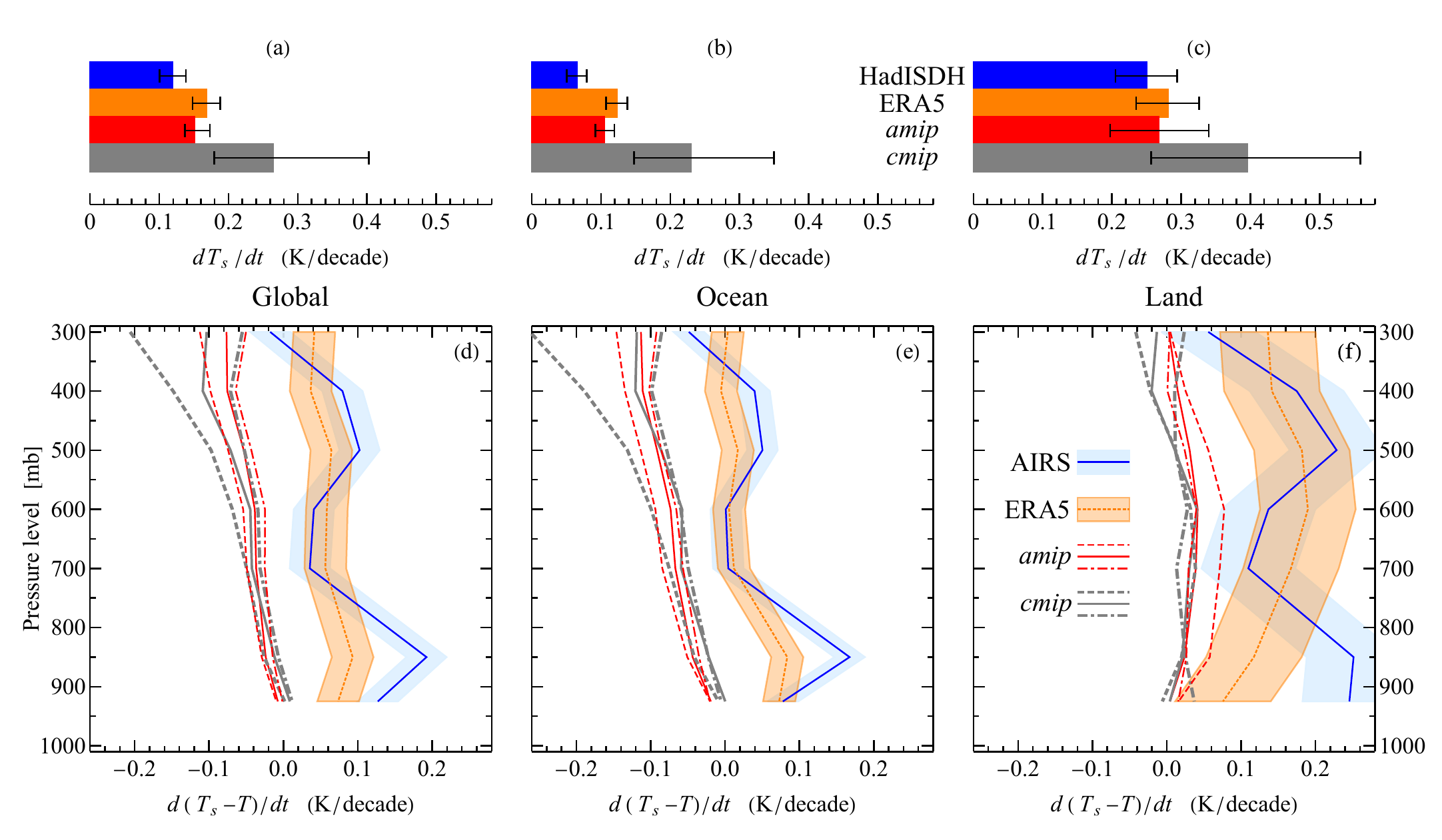}
\end{minipage}
\caption{
Mean trends of the surface temperature $T_s$ (a-c) and of the temperature difference $(T_s - T)$ between the surface and atmospheric pressure levels (d-f) for the planet as a whole (a,d), ocean (b,e) and land (c,f) over 1988--2014 in models (\textit{amip}, \textit{cmip}) versus observations (AIRS, HadISDH, ERA5). In (a-c), whiskers for HadISDH and ERA5 indicate $\pm$ one standard deviation, for \textit{amip} and \textit{cmip} -- the maximum and minimum values. Shading in (d-f) indicates $\pm$ one standard deviation. Dashed, solid and dash-dotted model curves in (d-f) were obtained, respectively, by using the maximum, median and minimum values of $dT_s/dt$ and $dT/dt$ in the model ensembles (\textit{amip} and \textit{cmip}). Data from Fig.~S4 of \citet{allan22}: AIRS, Atmospheric Infrared Sounder satellite data \citep{tian19}; HadISDH, The Met Office Hadley Centre homogenized and quality controlled, integrated sub-daily data set \citep{willett14}; ERA5, the fifth generation European Center for Medium-range Weather Forecasts global reanalysis \citep{hersbach20}; \textit{amip},  Atmospheric Model Intercomparison Project,  atmosphere-only simulations (without the ocean-atmosphere feedbacks in the climate system) \citep{gates99}; \textit{cmip}, Phase 6 of the Coupled Model Intercomparison Project (includes  \textit{amip} simulations as an integral part)
\citep{eyring16}. See \citet{allan22} for further details. In (d-f), $d(T_s - T)/dt$ for AIRS is calculated using HadISDH $dT_s/dt$.
Note that altitude $z(p)$ of a given pressure level $p$ increases slightly as the surface temperature grows, but for $p < 300$~mb it is a minor effect compared to the increase of
the temperature difference $T_s - T(p)$.
}
\label{temp}
\end{figure*}

In current models, it is assumed that as the planet warms, the temperature lapse rate should slightly diminish following moist adiabat \citep[the so-called lapse rate feedback,][]{sejas21}. This robust model feature is not, however, supported by observations (Fig.~\ref{temp}).
Satellite data are consistent with an increase in the lapse rate (Fig.~\ref{temp}). The temperature difference between the surface and the upper radiatve layer $z_e$ \citep[located between $500$ and $400$ mb,][]{benestad17} grows at approximately the same rate as the surface temperature itself. This effect is especially pronounced over land (Fig.~\ref{temp}c,f). This is consistent with a radiative forcing imposed by changing non-radiative fluxes, including those due to the land cover change (Fig.~\ref{ghe}d-f).

\section{Discussion and conclusions}

For the ecological audience it could be difficult to assess the credibility of our quantitative estimates, so we would like to emphasize two of the more unequivocal  points.  First,  global climate models do indicate that the regional loss of forest evapotranspiration leads to global warming. 
Eventhough the effect is small (Table~\ref{tab1}), it is of the opposite sign compared to the albedo-related cooling from deforestation that is invoked to argue that certain forests are not globally beneficial in the climate change context. Despite this obvious importance for the policy-relevant model outcomes, a conceptual description of how  evapotranspiration cools the Earth, and how its loss would lead to global warming,
is absent from the meteorological literature.  Conceptual understanding lacking, how can one independently assess whether the models get the effect right?

Second, global warming resulting from the loss of evapotranspiration is to be pronounced as an increase in the vertical lapse rate of air temperature. Due to the convective parameterization, global climate models keep this lapse rate roughly constant as the planet warms \citep{held06}.  However, this robust feature of global climate models does not agree with observations that accommodate a considerable increase in the temperature  difference between the surface and the upper radiative layer (Fig.~\ref{temp}). 

Policies based on the model outcomes that we have criticized are being shaped right now, and avoiding delays in their
re-evaluation is desirable. While the above arguments are percolating the meteorological literature, interested readers can approach their colleagues in the field of meteorology to see how they respond to the above two challenges, and thus get an indirect confirmation (or disproval) of our argumentation.

Our results highlight the importance of a {\it valid concept} put in the core of a model. The assumption of an {\it a priori} specified critical lapse rate
in the convective parameterization yields a negligible global transpirational cooling, which translates
into de-emphasizing the preservation of boreal forests. Concepts are powerful; incorrect concepts can be destructive. This brings us to the question, is there a {\it concept} that ecology could offer to put in the core of a global climate model, to adequately represent the biosphere?

From our perspective, it is the concept of environmental homeostasis, which is the capacity of natural ecosystems to compensate
for environmental disturbances and stabilize a favorable for life environment and climate \citep{lovelock74,vg95}. 
Recent studies discuss how the biotic control can be evident in the observed dynamics of the Earth{\textquoteright}s temperature
\citep{leggett20,leggett21,arnscheidt22}. When the information about how the natural ecosystem influences environment is lacking, the best guess could be to assume that they provide a stabilising feedback to the disturbance.

There was already a predicament in climate science that could have been facilitated by such an approach.
It was the {\textquotedblleft}missing sink{\textquotedblright} problem: when the rates of carbon accumulation in the atmosphere and the ocean became known with sufficient accuracy, it turned out that a signifcant part of fossil fuel emissions could not be accounted for.
The enigmatic {\textquotedblleft}missing sink{\textquotedblright} was later assigned to the terrestrial biota \citep{popkin15}.
While now it is commonly referred to as plant CO$_2$ fertilization, this is a response at the level of the ecological community as a whole: for there to be a net sink, as the plants synthesize more organic matter, heterotrophs must
refrain from decomposing it at a higher rate under the warming conditions \citep[cf.][]{wieder13}. 
Surprisingly, while the idea that {\it ecological succession proceeds in the direction of the ecosystem attaining increased
control of the environment and maximum protection from environmental perturbations} was dominant in ecology \citep{odum69}, a community{\textquoteright}s stabilizing response to the CO$_2$ disturbance was not predicted but rather opposed by ecologists on the basis that undisturbed ecosystems should have a closed matter cycle\footnote{This represents what can be called Odum{\textquoteright}s paradox, who thought that ecological succession culminates in ecosystem{\textquoteright}s maximum control of the environment \citep{odum69}. But if the ecosystem functions on the basis of closed matter cycles, its environmental impact (and, hence, environmental control) is zero by definition. The biotic regulation concept introduced the notion of {\it directed openness} of the matter cycles to compensate for environmental disturbances \citep{vg95}.} \citep{hampicke80,amthor95}. However, based on the premises of the biotic regulation concept \citep{vg95,ggm00}, and long before the missing sink was assigned to the terrestrial biota, \citet[][p.~946]{gorshkov86} predicted that the undisturbed ecosystems should perform a compensatory response to rising atmospheric CO$_2$ by elevating synthesis of carbohydrates.

Today, climate science faces a new challenge. Global climate models 
with an improved representation of clouds display a higher sensitivity of the Earth{\textquoteright}s climate to CO$_2$ doubling
than models with a poorer representation of clouds \citep{zelinka20,kuma22}. This implies more dire projections for future climate change, but also poses the problem of how to account for the past temperature changes that are not affected by the model improvements
and have been satisfactorily explained assuming a lower climate sensitivity.
The concept of the environmental homeostasis and the biotic regulation of the environment provide a possible solution:
the climate sensitivity may have been increasing with time -- reflecting the decline of natural ecosystems
and their global stabilising impact (Fig.~\ref{dist}).

Currently model uncertainties are assessed by comparing 
outputs from models developed by different research centers \citep{zelinka20}.
This provides a minimal uncertainty estimate, as the model development may follow universal principles
sharing both progress and errors. A distinct approach would be to attempt building a model
that departs significantly from the others in its core concept and see if such
a model can be plausibly tuned to competitively describe observations. Success of such a model
would force the range of model uncertainties to be extended.
As global climate models are currently being used to navigate our spacecraft Earth, with its multibillion crew, through
the storm of global climate disruption, such a stress test on their performance would not be superfluous.

Such an endeavour requires a plausible alternative concept, and we propose that  a global climate model built around the stabilising impact of natural ecosystems can
become such an alternative. This will require an interdisciplinary effort and an account of global transpirational cooling, the role of natural ecosystems in the long-distance moisture transport \citep{mg07,ent10,ellison2012,poveda14,molina19} and water cycle stabilization \citep{oconnor2021,baudena21,zemp17a}
and the distinct impact of ecosystems at different stages of ecological succession on the surface temperature and fire regime \citep[e.g.,][]{baker2019,aleinikov19,lindenmayer22} and the cloud cover \citep{cerasoli2021,duveiller2021}.
Living systems function on the basis of solar energy that under terrestrial conditions can be converted to useful work with a near $100\%$ efficiency. What processes are enacted with use of this energy, is determined by the genetic programs of all the organisms composing the ecological community. 
Randomly changing the species composition and morphological status of living organisms in the community --  for example, by replacing natural forest with a plantation with maximized productivity or by forcing the forest to remain in the early successional state \citep{kellett23} -- disturbs the flow of environmental information  and disrupts the ecosystem{\textquoteright}s capacity to respond to environmental disturbances.

While fundamental science is being advanced, the precautionary principle should be strictly applied. Any control system increases its feedback as the perturbation grows. Therefore, as the climate destabilisation deepens, the remaining natural ecosystems should be exerting an ever increasing compensatory impact per unit area.  In other words, the global climate price of losing a hectare of natural forest grows as the climate situation worsens. We call for an urgent global moratorium on the exploitation of the remaining natural ecosystems and a broad application of the {\it proforestation} strategy to allow them to restore to their full ecological and climate-regulating potential.

\section*{Acknowledgments}
The auhors are grateful to Jan Pokorn{\'y}, David Ellison, Ugo Bardi, Jon Schull and Zuzana Mulkerin for useful discussions. We thank Richard P. Allan for kindly supplying the data for trends in atmospheric and near surface temperature in numerical form. The work of A.M. Makarieva is partially funded by the Federal Ministry of Education and Research (BMBF) and the Free State of Bavaria under the Excellence Strategy of the Federal Government and the L{\"a}nder, as well as by the Technical University of Munich -- Institute for Advanced Study.


\begin{thebibliography}{71}
\providecommand{\natexlab}[1]{#1}
\providecommand{\url}[1]{{\tt #1}}
\providecommand{\urlprefix}{URL }
\expandafter\ifx\csname urlstyle\endcsname\relax
  \providecommand{\doi}[1]{https://doi.org/\discretionary{}{}{}#1}\else
  \providecommand{\doi}{https://doi.org/\discretionary{}{}{}\begingroup
  \urlstyle{rm}\Url}\fi

\bibitem[{Aleinikov(2019)}]{aleinikov19}
Aleinikov, A.: The fire history in pine forests of the plain area in the
  {Pechora-Ilych Nature Biosphere Reserve (Russia)} before 1942: {Possible}
  anthropogenic causes and long-term effects, Nat. Conserv. Res., 4, 21--34,
  \doi{10.24189/ncr.2019.033}, 2019.

\bibitem[{Alkama and Cescatti(2016)}]{alkama16}
Alkama, R. and Cescatti, A.: Biophysical climate impacts of recent changes in
  global forest cover, Science, 351, 600--604, \doi{10.1126/science.aac8083},
  2016.

\bibitem[{Allan et~al.(2022)Allan, Willett, John, and Trent}]{allan22}
Allan, R.~P., Willett, K.~M., John, V.~O., and Trent, T.: Global changes in
  water vapor 1979--2020, J. Geophys. Res.: Atmos., 127, e2022JD036\,728,
  \doi{10.1029/2022JD036728}, 2022.

\bibitem[{Amthor(1995)}]{amthor95}
Amthor, J.~S.: Terrestrial higher-plant response to increasing atmospheric
  {[CO$_{2}$]} in relation to the global carbon cycle, Glob. Change Biol., 1,
  243--274, \doi{10.1111/j.1365-2486.1995.tb00025.x}, 1995.

\bibitem[{Arnscheidt and Rothman(2022)}]{arnscheidt22}
Arnscheidt, C.~W. and Rothman, D.~H.: Presence or absence of stabilizing
  {Earth} system feedbacks on different time scales, Sci. Adv., 8, eadc9241,
  \doi{10.1126/sciadv.adc9241}, 2022.

\bibitem[{Arrhenius(1896)}]{arrhenius1896}
Arrhenius, S.: {XXXI. On} the influence of carbonic acid in the air upon the
  temperature of the ground, Lond. Edinb. Dublin Philos. Mag. J. Sci., 41,
  237--276, \doi{10.1080/14786449608620846}, 1896.

\bibitem[{Baker and Spracklen(2019)}]{baker2019}
Baker, J. C.~A. and Spracklen, D.~V.: Climate benefits of intact {Amazon}
  forests and the biophysical consequences of disturbance, Front. For. Glob.
  Change, 2, \doi{10.3389/ffgc.2019.00047}, 2019.

\bibitem[{Bates(1999)}]{bates99}
Bates, J.~R.: A dynamical stabilizer in the climate system: {A} mechanism
  suggested by a simple model, Tellus A: Dyn. Meteorol. Oceanogr., 51,
  349--372, \doi{10.3402/tellusa.v51i3.13458}, 1999.

\bibitem[{Bates(2003)}]{bates03}
Bates, J.~R.: On climate stability, climate sensitivity and the dynamics of the
  enhanced greenhouse effect, DCESS REPORT, 3, 1--38, 2003.

\bibitem[{Baudena et~al.(2021)Baudena, Tuinenburg, Ferdinand, and
  Staal}]{baudena21}
Baudena, M., Tuinenburg, O.~A., Ferdinand, P.~A., and Staal, A.: Effects of
  land-use change in the {Amazon} on precipitation are likely underestimated,
  Glob. Change Biol., 27, 5580--5587, \doi{10.1111/gcb.15810}, 2021.

\bibitem[{Benestad(2017)}]{benestad17}
Benestad, R.~E.: A mental picture of the greenhouse effect, Theor. Appl.
  Climatol., 128, 679--688, \doi{10.1007/s00704-016-1732-y}, 2017.

\bibitem[{Caballero(2001)}]{caballero01}
Caballero, R.: Surface wind, subcloud humidity and the stability of the
  tropical climate, Tellus A: Dyn. Meteorol. Oceanogr., 53, 513--525,
  \doi{10.3402/tellusa.v53i4.12224}, 2001.

\bibitem[{Cerasoli et~al.(2021)Cerasoli, Yin, and Porporato}]{cerasoli2021}
Cerasoli, S., Yin, J., and Porporato, A.: Cloud cooling effects of
  afforestation and reforestation at midlatitudes, Proc. Natl. Acad. Sci. USA,
  118, \doi{10.1073/pnas.2026241118}, 2021.

\bibitem[{Chaplin-Kramer et~al.(2022)Chaplin-Kramer, Neugarten, Sharp, Collins,
  Polasky, Hole, Schuster, Strimas-Mackey, Mulligan, Brandon, Diaz,
  Fluet-Chouinard, Gorenflo, Johnson, Kennedy, Keys, Longley-Wood, McIntyre,
  Noon, Pascual, Reidy~Liermann, Roehrdanz, Schmidt-Traub, Shaw, Spalding,
  Turner, van Soesbergen, and Watson}]{chaplin-kramer22}
Chaplin-Kramer, R., Neugarten, R.~A., Sharp, R.~P., Collins, P.~M., Polasky,
  S., Hole, D., Schuster, R., Strimas-Mackey, M., Mulligan, M., Brandon, C.,
  Diaz, S., Fluet-Chouinard, E., Gorenflo, L.~J., Johnson, J.~A., Kennedy,
  C.~M., Keys, P.~W., Longley-Wood, K., McIntyre, P.~B., Noon, M., Pascual, U.,
  Reidy~Liermann, C., Roehrdanz, P.~R., Schmidt-Traub, G., Shaw, M.~R.,
  Spalding, M., Turner, W.~R., van Soesbergen, A., and Watson, R.~A.: Mapping
  the planet{\textquoteright}s critical natural assets, Nat. Ecol. Evol.,
  \doi{10.1038/s41559-022-01934-5}, 2022.

\bibitem[{Davin and de~Noblet-Ducoudr\'{e}(2010)}]{davin10}
Davin, E.~L. and de~Noblet-Ducoudr\'{e}, N.: Climatic impact of global-scale
  deforestation: {Radiative} versus nonradiative processes, J. Clim., 23,
  97--112, \doi{10.1175/2009JCLI3102.1}, 2010.

\bibitem[{De~Hertog et~al.(2022)De~Hertog, Havermann, Vanderkelen, Guo, Luo,
  Manola, Coumou, Davin, Duveiller, Lejeune, Pongratz, Schleussner,
  Seneviratne, and Thiery}]{hertog22}
De~Hertog, S.~J., Havermann, F., Vanderkelen, I., Guo, S., Luo, F., Manola, I.,
  Coumou, D., Davin, E.~L., Duveiller, G., Lejeune, Q., Pongratz, J.,
  Schleussner, C.-F., Seneviratne, S.~I., and Thiery, W.: The biogeophysical
  effects of idealized land cover and land management changes in {Earth} system
  models, Earth Syst. Dynam., 13, 1305--1350, \doi{10.5194/esd-13-1305-2022},
  2022.

\bibitem[{Duveiller et~al.(2021)Duveiller, Filipponi, Ceglar, Bojanowski,
  Alkama, and Cescatti}]{duveiller2021}
Duveiller, G., Filipponi, F., Ceglar, A., Bojanowski, J., Alkama, R., and
  Cescatti, A.: Revealing the widespread potential of forests to increase low
  level cloud cover, Nat. Commun., 12, 4337, \doi{10.1038/s41467-021-24551-5},
  2021.

\bibitem[{Ellison et~al.(2012)Ellison, N.~Futter, and Bishop}]{ellison2012}
Ellison, D., N.~Futter, M., and Bishop, K.: On the forest cover--water yield
  debate: from demand- to supply-side thinking, Glob. Change Biol., 18,
  806--820, \doi{10.1111/j.1365-2486.2011.02589.x}, 2012.

\bibitem[{Ellison et~al.(2017)Ellison, Morris, Locatelli, Sheil, Cohen,
  Murdiyarso, Gutierrez, van Noordwijk, Creed, Pokorny, Gaveau, Spracklen,
  Tobella, Ilstedt, Teuling, Gebrehiwot, Sands, Muys, Verbist, Springgay,
  Sugandi, and Sullivan}]{ellison2017}
Ellison, D., Morris, C.~E., Locatelli, B., Sheil, D., Cohen, J., Murdiyarso,
  D., Gutierrez, V., van Noordwijk, M., Creed, I.~F., Pokorny, J., Gaveau, D.,
  Spracklen, D.~V., Tobella, A.~B., Ilstedt, U., Teuling, A.~J., Gebrehiwot,
  S.~G., Sands, D.~C., Muys, B., Verbist, B., Springgay, E., Sugandi, Y., and
  Sullivan, C.~A.: Trees, forests and water: {Cool} insights for a hot world,
  Glob. Environ. Change, 43, 51--61, \doi{10.1016/j.gloenvcha.2017.01.002},
  2017.

\bibitem[{Eyring et~al.(2016)Eyring, Bony, Meehl, Senior, Stevens, Stouffer,
  and Taylor}]{eyring16}
Eyring, V., Bony, S., Meehl, G.~A., Senior, C.~A., Stevens, B., Stouffer,
  R.~J., and Taylor, K.~E.: Overview of the {Coupled Model Intercomparison
  Project Phase 6 (CMIP6)} experimental design and organization, Geosci. Model
  Dev., 9, 1937--1958, \doi{10.5194/gmd-9-1937-2016}, 2016.

\bibitem[{Gates et~al.(1999)Gates, Boyle, Covey, Dease, Doutriaux, Drach,
  Fiorino, Gleckler, Hnilo, Marlais, Phillips, Potter, Santer, Sperber, Taylor,
  and Williams}]{gates99}
Gates, W.~L., Boyle, J.~S., Covey, C., Dease, C.~G., Doutriaux, C.~M., Drach,
  R.~S., Fiorino, M., Gleckler, P.~J., Hnilo, J.~J., Marlais, S.~M., Phillips,
  T.~J., Potter, G.~L., Santer, B.~D., Sperber, K.~R., Taylor, K.~E., and
  Williams, D.~N.: An overview of the results of the {Atmospheric Model
  Intercomparison Project (AMIP I)}, Bull. Amer. Meteor. Soc., 80, 29--56,
  \doi{10.1175/1520-0477(1999)080<0029:AOOTRO>2.0.CO;2}, 1999.

\bibitem[{Goody(2003)}]{goody03}
Goody, R.: On the mechanical efficiency of deep, tropical convection, J. Atmos.
  Sci., 60, 2827--2832, \doi{10.1175/1520-0469(2003)060<2827:OTMEOD>2.0.CO;2},
  2003.

\bibitem[{Gorshkov et~al.(2002)Gorshkov, Makarieva, and Pujol}]{gorshkov02}
Gorshkov, V., Makarieva, A., and Pujol, T.: Radiative-convective processes and
  changes of the flux of thermal radiation into space with increasing optical
  thickness of the atmosphere, pp. 499--525, PNPI RAS, Gatchina, St.
  Petersburg, Russia, {Proceedings of the XXXVI Winter School of Petersburg
  Nuclear Physics Institute (Nuclear and Particle Physics), February 25 --
  March 3}, 2002.

\bibitem[{Gorshkov(1986)}]{gorshkov86}
Gorshkov, V.~G.: Atmospheric disturbance of the carbon cycle: {Impact} upon the
  biosphere, Nuov. Cim. C, 9, 937--952, \doi{10.1007/BF02891905}, 1986.

\bibitem[{Gorshkov(1995)}]{vg95}
Gorshkov, V.~G.: Physical and biological bases of life stability. {Man, Biota,
  Environment}, Springer, Berlin, Heidelberg, \doi{10.1007/978-3-642-85001-1},
  1995.

\bibitem[{Gorshkov et~al.(2000)Gorshkov, Gorshkov, and Makarieva}]{ggm00}
Gorshkov, V.~G., Gorshkov, V.~V., and Makarieva, A.~M.: Biotic regulation of
  the environment: {Key} issue of global change, Springer, Berlin, 2000.

\bibitem[{Hagemann et~al.(2011)Hagemann, Chen, Haerter, Heinke, Gerten, and
  Piani}]{hagemann11}
Hagemann, S., Chen, C., Haerter, J.~O., Heinke, J., Gerten, D., and Piani, C.:
  Impact of a statistical bias correction on the projected hydrological changes
  obtained from three {GCMs} and two hydrology models, J. Hydrometeor., 12,
  556--578, \doi{10.1175/2011JHM1336.1}, 2011.

\bibitem[{Hampicke(1980)}]{hampicke80}
Hampicke, U.: The effect of the atmosphere-biosphere exchange on the global
  carbon cycle, Experientia, 36, 776--781, \doi{10.1007/BF01978577}, 1980.

\bibitem[{Held and Soden(2006)}]{held06}
Held, I.~M. and Soden, B.~J.: Robust responses of the hydrological cycle to
  global warming, J. Clim., 19, 5686--5699, \doi{10.1175/JCLI3990.1}, 2006.

\bibitem[{Hersbach et~al.(2020)Hersbach, Bell, Berrisford, Hirahara,
  Hor{\'a}nyi, {Mu\~{n}oz-Sabater}, Nicolas, Peubey, Radu, Schepers, Simmons,
  Soci, Abdalla, Abellan, Balsamo, Bechtold, Biavati, Bidlot, Bonavita,
  De~Chiara, Dahlgren, Dee, Diamantakis, Dragani, Flemming, Forbes, Fuentes,
  Geer, Haimberger, Healy, Hogan, H{\'o}lm, Janiskov{\'a}, Keeley, Laloyaux,
  Lopez, Lupu, Radnoti, de~Rosnay, Rozum, Vamborg, Villaume, and
  Th{\'e}paut}]{hersbach20}
Hersbach, H., Bell, B., Berrisford, P., Hirahara, S., Hor{\'a}nyi, A.,
  {Mu\~{n}oz-Sabater}, J., Nicolas, J., Peubey, C., Radu, R., Schepers, D.,
  Simmons, A., Soci, C., Abdalla, S., Abellan, X., Balsamo, G., Bechtold, P.,
  Biavati, G., Bidlot, J., Bonavita, M., De~Chiara, G., Dahlgren, P., Dee, D.,
  Diamantakis, M., Dragani, R., Flemming, J., Forbes, R., Fuentes, M., Geer,
  A., Haimberger, L., Healy, S., Hogan, R.~J., H{\'o}lm, E., Janiskov{\'a}, M.,
  Keeley, S., Laloyaux, P., Lopez, P., Lupu, C., Radnoti, G., de~Rosnay, P.,
  Rozum, I., Vamborg, F., Villaume, S., and Th{\'e}paut, J.-N.: The {ERA5}
  global reanalysis, Quart. J. Roy. Meteor. Soc., 146, 1999--2049,
  \doi{10.1002/qj.3803}, 2020.

\bibitem[{Hesslerov{\'a} et~al.(2018)Hesslerov{\'a}, Huryna, Pokorn{\'y}, and
  Proch{\'a}zka}]{hesslerova18}
Hesslerov{\'a}, P., Huryna, H., Pokorn{\'y}, J., and Proch{\'a}zka, J.: The
  effect of forest disturbance on landscape temperature, Ecol. Eng., 120,
  345--354, \doi{10.1016/j.ecoleng.2018.06.011}, 2018.

\bibitem[{Heus and Jonker(2008)}]{heus08}
Heus, T. and Jonker, H. J.~J.: Subsiding shells around shallow cumulus clouds,
  J. Atmos. Sci., 65, 1003--1018, \doi{10.1175/2007JAS2322.1}, 2008.

\bibitem[{Hurtt et~al.(2020)Hurtt, Chini, Sahajpal, Frolking, Bodirsky, Calvin,
  Doelman, Fisk, Fujimori, Klein~Goldewijk, Hasegawa, Havlik, Heinimann,
  Humpen\"{o}der, Jungclaus, Kaplan, Kennedy, Krisztin, Lawrence, Lawrence, Ma,
  Mertz, Pongratz, Popp, Poulter, Riahi, Shevliakova, Stehfest, Thornton,
  Tubiello, van Vuuren, and Zhang}]{hurtt20}
Hurtt, G.~C., Chini, L., Sahajpal, R., Frolking, S., Bodirsky, B.~L., Calvin,
  K., Doelman, J.~C., Fisk, J., Fujimori, S., Klein~Goldewijk, K., Hasegawa,
  T., Havlik, P., Heinimann, A., Humpen\"{o}der, F., Jungclaus, J., Kaplan,
  J.~O., Kennedy, J., Krisztin, T., Lawrence, D., Lawrence, P., Ma, L., Mertz,
  O., Pongratz, J., Popp, A., Poulter, B., Riahi, K., Shevliakova, E.,
  Stehfest, E., Thornton, P., Tubiello, F.~N., van Vuuren, D.~P., and Zhang,
  X.: Harmonization of global land use change and management for the period
  850--2100 {(LUH2) for CMIP6}, Geosci. Model Dev., 13, 5425--5464,
  \doi{10.5194/gmd-13-5425-2020}, 2020.

\bibitem[{Huryna and Pokorn{\'y}(2016)}]{hurina16}
Huryna, H. and Pokorn{\'y}, J.: The role of water and vegetation in the
  distribution of solar energy and local climate: {A} review, Folia Geobot.,
  51, 191--208, \doi{10.1007/s12224-016-9261-0}, 2016.

\bibitem[{IPBES(2019)}]{ipbes19}
IPBES: {Global assessment report on biodiversity and ecosystem services of the
  Intergovernmental Science-Policy Platform on Biodiversity and Ecosystem
  Services}, IPBES secretariat, Bonn, Germany, \doi{10.5281/zenodo.6417333},
  1148 pages, 2019.

\bibitem[{IPCC(2021)}]{ipcc21}
IPCC: Climate Change 2021: The Physical Science Basis. Contribution of Working
  Group I to the Sixth Assessment Report of the Intergovernmental Panel on
  Climate Change, Cambridge University Press, Cambridge, United Kingdom and New
  York, NY, USA, \doi{10.1017/9781009157896}, in press, 2021.

\bibitem[{Jia et~al.(2019)Jia, Shevliakova, Artaxo, De~Noblet-Ducoudr{\'e},
  Houghton, House, Kitajima, Lennard, Popp, Sirin, Sukumar, and
  Verchot}]{ipcc19}
Jia, G., Shevliakova, E., Artaxo, P., De~Noblet-Ducoudr{\'e}, N., Houghton, R.,
  House, J., Kitajima, K., Lennard, C., Popp, A., Sirin, A., Sukumar, R., and
  Verchot, L.: Chapter 2. {Land-climate} interactions, in: Climate Change and
  Land: an {IPCC} special report on climate change, desertification, land
  degradation, sustainable land management, food security, and greenhouse gas
  fluxes in terrestrial ecosystems, edited by Shukla, P.~R., Skea, J.,
  Calvo~Buendia, E., Masson-Delmotte, V., P{\"{o}}rtner, H.-O., Roberts, D.~C.,
  Zhai, P., Slade, R., Connors, S., van Diemen, R., Ferrat, M., Haughey, E.,
  Luz, S., Neogi, S., Pathak, M., Petzold, J., Portugal~Pereira, J., Vyas, P.,
  Huntley, E., Kissick, K., Belkacemi, M., and Malley, J.,
  \urlprefix\url{https://www.ipcc.ch/srccl/cite-report/}, in press, 2019.

\bibitem[{Katzwinkel et~al.(2014)Katzwinkel, Siebert, Heus, and
  Shaw}]{katzwinkel14}
Katzwinkel, J., Siebert, H., Heus, T., and Shaw, R.~A.: Measurements of
  turbulent mixing and subsiding shells in trade wind cumuli, J. Atmos. Sci.,
  71, 2810--2822, \doi{10.1175/JAS-D-13-0222.1}, 2014.

\bibitem[{Kellett et~al.(2023)Kellett, Maloof, Masino, Frelich, Faison, Brosi,
  and Foster}]{kellett23}
Kellett, M.~J., Maloof, J.~E., Masino, S.~A., Frelich, L.~E., Faison, E.~K.,
  Brosi, S.~L., and Foster, D.~R.: Forest-clearing to create early-successional
  habitats: {Questionable} benefits, significant costs, Front. For. Glob.
  Change, 5, \doi{10.3389/ffgc.2022.1073677}, 2023.

\bibitem[{Kuma et~al.(2022)Kuma, Bender, Schuddeboom, McDonald, and
  Seland}]{kuma22}
Kuma, P., Bender, F. A.-M., Schuddeboom, A., McDonald, A.~J., and Seland,
  {\O}.: Machine learning of cloud types shows higher climate sensitivity is
  associated with lower cloud biases, Atmos. Chem. Phys. Discuss., 2022, 1--32,
  \doi{10.5194/acp-2022-184}, [preprint], 2022.

\bibitem[{Leggett and Ball(2020)}]{leggett20}
Leggett, L. M.~W. and Ball, D.~A.: Observational evidence that a feedback
  control system with proportional-integral-derivative characteristics is
  operating on atmospheric surface temperature at global scale, Tellus A: Dyn.
  Meteorol. Oceanogr., 72, 1--14, \doi{10.1080/16000870.2020.1717268}, 2020.

\bibitem[{Leggett and Ball(2021)}]{leggett21}
Leggett, L. M.~W. and Ball, D.~A.: Empirical evidence for a global atmospheric
  temperature control system: physical structure, Tellus A: Dyn. Meteorol.
  Oceanogr., 73, 1--24, \doi{10.1080/16000870.2021.1926123}, 2021.

\bibitem[{Lejeune et~al.(2017)Lejeune, Seneviratne, and Davin}]{lejeune17}
Lejeune, Q., Seneviratne, S.~I., and Davin, E.~L.: Historical land-cover change
  impacts on climate: {Comparative} assessment of {LUCID} and {CMIP5}
  multimodel experiments, J. Clim., 30, 1439--1459,
  \doi{10.1175/JCLI-D-16-0213.1}, 2017.

\bibitem[{Lindenmayer et~al.(2022)Lindenmayer, Bowd, Taylor, and
  Likens}]{lindenmayer22}
Lindenmayer, D.~B., Bowd, E.~J., Taylor, C., and Likens, G.~E.: The
  interactions among fire, logging, and climate change have sprung a landscape
  trap in {Victoria}{\textquoteright}s montane ash forests, Plant Ecol., 223,
  733--749, \doi{10.1007/s11258-021-01217-2}, 2022.

\bibitem[{Lovelock and Margulis(1974)}]{lovelock74}
Lovelock, J. and Margulis, L.: Atmospheric homeostasis by and for the
  biosphere: the gaia hypothesis, Tellus, 26, 2--10,
  \doi{10.3402/tellusa.v26i1-2.9731}, 1974.

\bibitem[{Makarieva and Gorshkov(2001)}]{makarieva01}
Makarieva, A. and Gorshkov, V.: The greenhouse effect and the stability of the
  global mean surface temperature, Dokl. Earth Sci., 377, 210--214, 2001.

\bibitem[{Makarieva and Gorshkov(2007)}]{mg07}
Makarieva, A.~M. and Gorshkov, V.~G.: Biotic pump of atmospheric moisture as
  driver of the hydrological cycle on land, Hydrol. Earth Syst. Sci., 11,
  1013--1033, \doi{10.5194/hess-11-1013-2007}, 2007.

\bibitem[{Marengo(2006)}]{ma06}
Marengo, J.~A.: On the hydrological cycle of the {Amazon} basin: {A} historical
  review and current state-of-the-art, Rev. Bras. Meteorol., 21, 1--19, 2006.

\bibitem[{Marvel et~al.(2013)Marvel, Kravitz, and Caldeira}]{marvel13}
Marvel, K., Kravitz, B., and Caldeira, K.: Geophysical limits to global wind
  power, Nat. Clim. Change, 3, 118--121, \doi{10.1038/nclimate1683}, 2013.

\bibitem[{Molina et~al.(2019)Molina, Salazar, Mart{\'{\i}}nez, Villegas, and
  Arias}]{molina19}
Molina, R.~D., Salazar, J.~F., Mart{\'{\i}}nez, J.~A., Villegas, J.~C., and
  Arias, P.~A.: Forest-induced exponential growth of precipitation along
  climatological wind streamlines over the {Amazon}, J. Geophys. Res.: Atmos.,
  124, 2589--2599, \doi{10.1029/2018JD029534}, 2019.

\bibitem[{Moomaw et~al.(2019)Moomaw, Masino, and Faison}]{moomaw2019}
Moomaw, W.~R., Masino, S.~A., and Faison, E.~K.: Intact forests in the {United
  States: Proforestation} mitigates climate change and serves the greatest
  good, Front. For. Glob. Change, 2, \doi{10.3389/ffgc.2019.00027}, 2019.

\bibitem[{O{\textquoteright}Connor et~al.(2021)O{\textquoteright}Connor,
  Dekker, Staal, Tuinenburg, Rebel, and Santos}]{oconnor2021}
O{\textquoteright}Connor, J.~C., Dekker, S.~C., Staal, A., Tuinenburg, O.~A.,
  Rebel, K.~T., and Santos, M.~J.: Forests buffer against variations in
  precipitation, Glob. Change Biol., 27, 4686--4696, \doi{10.1111/gcb.15763},
  2021.

\bibitem[{Odum(1969)}]{odum69}
Odum, E.~P.: The strategy of ecosystem development: {An} understanding of
  ecological succession provides a basis for resolving man{\textquoteright}s
  conflict with nature, Science, 164, 262--270,
  \doi{10.1126/science.164.3877.262}, 1969.

\bibitem[{Pearce(2022)}]{pearce22}
Pearce, F.: The forest forecast, Science, 376, 788--791,
  \doi{10.1126/science.adc9867}, 2022.

\bibitem[{Popkin(2015)}]{popkin15}
Popkin, G.: The hunt for the world{\textquoteright}s missing carbon, Nature,
  523, 20--22, \doi{10.1038/523020a}, 2015.

\bibitem[{Poveda et~al.(2014)Poveda, Jaramillo, and Vallejo}]{poveda14}
Poveda, G., Jaramillo, L., and Vallejo, L.~F.: Seasonal precipitation patterns
  along pathways of {South American} low-level jets and aerial rivers, Water
  Resour. Res., 50, 98--118, \doi{10.1002/2013WR014087}, 2014.

\bibitem[{Ramanathan and Coakley~Jr.(1978)}]{ramanathan78}
Ramanathan, V. and Coakley~Jr., J.~A.: Climate modeling through
  radiative-convective models, Rev. Geophys., 16, 465--489,
  \doi{10.1029/RG016i004p00465}, 1978.

\bibitem[{Schepaschenko et~al.(2021)Schepaschenko, Moltchanova, Fedorov,
  Karminov, Ontikov, Santoro, See, Kositsyn, Shvidenko, Romanovskaya, Korotkov,
  Lesiv, Bartalev, Fritz, Shchepashchenko, and Kraxner}]{schepaschenko21}
Schepaschenko, D., Moltchanova, E., Fedorov, S., Karminov, V., Ontikov, P.,
  Santoro, M., See, L., Kositsyn, V., Shvidenko, A., Romanovskaya, A.,
  Korotkov, V., Lesiv, M., Bartalev, S., Fritz, S., Shchepashchenko, M., and
  Kraxner, F.: Russian forest sequesters substantially more carbon than
  previously reported, Sci. Rep., 11, 12\,825,
  \doi{10.1038/s41598-021-92152-9}, 2021.

\bibitem[{Sejas et~al.(2021)Sejas, Hu, Cai, and Fan}]{sejas21}
Sejas, S., Hu, X., Cai, M., and Fan, H.: Understanding the differences between
  {TOA} and surface energy budget attributions of surface warming, Front. Earth
  Sci., 9, \doi{10.3389/feart.2021.725816}, 2021.

\bibitem[{Seymour et~al.(2022{\natexlab{a}})Seymour, Wolosin, and
  Gray}]{seymour22a}
Seymour, F., Wolosin, M., and Gray, E.: Policies underestimate
  forests{\textquoteright} full effect on the climate,
  \url{https://www.wri.org/insights/how-forests-affect-climate}, {October 23},
  2022{\natexlab{a}}.

\bibitem[{Seymour et~al.(2022{\natexlab{b}})Seymour, Wolosin, and
  Gray}]{seymour22b}
Seymour, F., Wolosin, M., and Gray, E.: Not just carbon: {Capturing} all the
  benefits of forests for stabilizing the climate from local to global scales,
  Report, {Washington, DC}, \doi{10.46830/wrirpt.19.00004}, 2022{\natexlab{b}}.

\bibitem[{Snyder et~al.(2004)Snyder, Delire, and Foley}]{snyder04}
Snyder, P.~K., Delire, C., and Foley, J.~A.: Evaluating the influence of
  different vegetation biomes on the global climate, Clim. Dynam., 23,
  279--302, \doi{10.1007/s00382-004-0430-0}, 2004.

\bibitem[{Sohail et~al.(2022)Sohail, Zika, Irving, and Church}]{sohail22}
Sohail, T., Zika, J.~D., Irving, D.~B., and Church, J.~A.: Observed poleward
  freshwater transport since 1970, Nature, 602, 617--622,
  \doi{10.1038/s41586-021-04370-w}, 2022.

\bibitem[{Tian et~al.(2019)Tian, Fetzer, and Manning}]{tian19}
Tian, B., Fetzer, E.~J., and Manning, E.~M.: {The Atmospheric Infrared Sounder
  Obs4MIPs Version 2} data set, Earth Space Sci., 6, 324--333,
  \doi{10.1029/2018EA000508}, 2019.

\bibitem[{Trenberth et~al.(2009)Trenberth, Fasullo, and Kiehl}]{trenberth09}
Trenberth, K.~E., Fasullo, J.~T., and Kiehl, J.: Earth{\textquoteright}s global
  energy budget, Bull. Amer. Meteor. Soc., 90, 311--324,
  \doi{10.1175/2008BAMS2634.1}, 2009.

\bibitem[{van~der Ent et~al.(2010)van~der Ent, Savenije, Schaefli, and
  Steele-Dunne}]{ent10}
van~der Ent, R.~J., Savenije, H. H.~G., Schaefli, B., and Steele-Dunne, S.~C.:
  Origin and fate of atmospheric moisture over continents, Water Resour. Res.,
  46, W09\,525, \doi{10.1029/2010WR009127}, 2010.

\bibitem[{Wieder et~al.(2013)Wieder, Bonan, and Allison}]{wieder13}
Wieder, W.~R., Bonan, G.~B., and Allison, S.~D.: Global soil carbon projections
  are improved by modelling microbial processes, Nat. Clim. Change, 3,
  909--912, \doi{10.1038/nclimate1951}, 2013.

\bibitem[{Wilhere(2021)}]{wilhere21}
Wilhere, G.~F.: A {Paris-like} agreement for biodiversity needs {IPCC-like}
  science, Glob. Ecol. Conserv., 28, \doi{10.1016/j.gecco.2021.e01617}, 2021.

\bibitem[{Willett et~al.(2014)Willett, Dunn, Thorne, Bell, de~Podesta, Parker,
  Jones, and Williams~Jr.}]{willett14}
Willett, K.~M., Dunn, R. J.~H., Thorne, P.~W., Bell, S., de~Podesta, M.,
  Parker, D.~E., Jones, P.~D., and Williams~Jr., C.~N.: {HadISDH} land surface
  multi-variable humidity and temperature record for climate monitoring, Clim.
  Past, 10, 1983--2006, \doi{10.5194/cp-10-1983-2014}, 2014.

\bibitem[{Zelinka et~al.(2020)Zelinka, Myers, McCoy, Po-Chedley, Caldwell,
  Ceppi, Klein, and Taylor}]{zelinka20}
Zelinka, M.~D., Myers, T.~A., McCoy, D.~T., Po-Chedley, S., Caldwell, P.~M.,
  Ceppi, P., Klein, S.~A., and Taylor, K.~E.: Causes of higher climate
  sensitivity in {CMIP6} models, Geophys. Res. Lett., 47, e2019GL085\,782,
  \doi{https://doi.org/10.1029/2019GL085782}, 2020.

\bibitem[{Zemp et~al.(2017)Zemp, Schleussner, Barbosa, and Rammig}]{zemp17a}
Zemp, D.~C., Schleussner, C.-F., Barbosa, H. M.~J., and Rammig, A.:
  Deforestation effects on {Amazon} forest resilience, Geophys. Res. Lett., 44,
  6182--6190, \doi{10.1002/2017GL072955}, 2017.

\end{thebibliography}


\end{document}